\documentclass[reqno]{amsart}

\usepackage[hyphens]{url}
\usepackage[style=numeric,sorting=none,backend=biber,maxbibnames=99]{biblatex}
\usepackage{tiz-note}
\bibliography{main}
\usepackage{tiz-paper}
\usepackage{fontawesome}

\definecolor{AyaFn}{HTML}{00627a}
\definecolor{AyaConstructor}{HTML}{067d17}
\definecolor{AyaStruct}{HTML}{00627a}
\definecolor{AyaGeneralized}{HTML}{00627a}
\definecolor{AyaData}{HTML}{00627a}
\definecolor{AyaPrimitive}{HTML}{00627a}
\definecolor{AyaKeyword}{HTML}{0033b3}
\definecolor{AyaComment}{HTML}{8c8c8c}
\definecolor{AyaField}{HTML}{871094}
\newcommand\AyaFn[1]{\textcolor{AyaFn}{#1}}
\newcommand\AyaConstructor[1]{\textcolor{AyaConstructor}{#1}}

\newcommand\AyaData[1]{\textcolor{AyaData}{#1}}

\newcommand\AyaKeyword[1]{\textcolor{AyaKeyword}{#1}}
\newcommand\AyaLocalVar[1]{\textit{#1}}

\newcommand{\inS}[1]{\textsf{inS}(#1)}
\newcommand{\outS}[1]{\textsf{outS}(#1)}

\pgfdeclarelayer{frontmost}
\pgfsetlayers{main,frontmost}
\usetikzlibrary{patterns}

\tikzset{
  carlo-axes/.style =
  {
    y = {(0,-1)},
    z = {(-0.6,0.6)}
  } ,
  shorten <>/.style =
  {
    shorten >=#1 , shorten <=#1
  } ,
  equals arrow/.style =
  {
    arrows = - ,
    double equal sign distance ,
  } ,
}

\usepackage[hyphenbreaks]{breakurl}

\begin{document}

\title{Three non-cubical applications of extension types}

\author{Tesla Zhang}
\address{Carnegie Mellon University}
\email{teslaz@cmu.edu}
\date\today

\begin{abstract}
The development of cubical type theory inspired the idea of ``extension types''
which has been found to have applications in other type theories that are unrelated
to homotopy type theory or cubical type theory.

This article describes these applications, including on records, metaprogramming,
controlling unfolding, and some more exotic ones.
\end{abstract}
\maketitle
\tableofcontents
This article is intended to introduce a simple version of \textit{extension types}
along with PL applications that are unrelated to its original motivation,
which is related to homotopy type theory (HoTT)~\cite{HoTTBook}.
This is not saying HoTT is not PL, but rather that
the context of HoTT is big, and it's not necessary for understanding
the idea of extension types.

We will not be talking too much about type theories with shapes,
simplicial type theory~\cite{InfCat}, or cubical type theory~\cite{CCHM,CHM,CCTT}.
However, we give readers who are familiar with cubical a hint of how
extension types are related to the cubical path type.

Most of these ideas are neither new nor mine.
Some of them are from personal communications with Reed Mullanix,
Andr\'as Kov\'acs, and Harrison Grodin.
\Cref{sub:defproj} is my own work inspired from Reed's ideas,
Joseph Hua contributed the idea in \cref{rem:extent} through personal communication,
Steve Awodey commented on \cref{rem:extt} and suggested the use of $▷$
during the presentation of this article in the HoTT seminar.

\section{Applications of extension types}\label{sec:problems}
Before getting into extension types, we first introduce some desirable features
in dependent type theories. Some can be implemented without extension types,
but extension types have their own benefit that their metatheory is fairly well understood.

\subsection{Controlling unfolding}
There are times we don't want to unfold definitions arbitrarily when programming in dependent type theory.
Imagine the following scenarios, which are also discussed in~\cite{CU}
(``cluttered proof states and unreadable error messages''):
\begin{itemize}
\item We are to prove $f(u) = f(v)$ where $f$ expands to a gigantic term,
  what we really need to know is that we need to show $u = v$, but the type checker is free to
  just unfold $f$ since $\beta$-reduction is an equality,
  which would make it uneasy to see what is really going on.
  Function definitions exist for reasons!
\item We wrote a library, and it's used by a downstream codebase, and we want to change the library.
  In a non-dependently typed language, we can guarantee backward compatibility by not changing the interface,
  say, the types. But in our case, this is no longer true,
  because definitions are part of the interfaces,
  and if the downstream relies on certain functions to unfold in certain ways, then we can't change that.
  But how do we know?
\end{itemize}
That is why controlling unfolding is important. What gets proposed is that we now make explicit
in each definition that, during the type checking of this defining,
what other definitions we intend to unfold, and by default we don't unfold anything.
There are certain things we wish to guarantee:
\begin{itemize}
\item Unfolding is \textit{transitive}. If function $f$ unfolds $g$ while $g$ unfolds $h$,
  then in $f$, we also unfold $h$.
\item \textit{Normalization} and \textit{canonicity}. Since we're touching judgmental equality now,
  normalization in this case becomes a bit more complicated.
  Sometimes we have a definition that is not unfolded, but it's only because we do not want it to,
  while traditional normalization needs us to unfold everything.
  There are things we need to change here --- either how we define normalization or how we think about definitions.
\end{itemize}

\subsection{Partially applied records with definitional projection}\label{sub:motive-defproj}
This section talks about a language feature called \textit{anonymous extensions}
in the Arend prover\footnote{\url{https://arend-lang.github.io/about/arend-features\#anonymous-extensions}}.

Suppose we have the following (dependent) record\footnote{For those who are curious,
this is part of a \textit{precategory} in the context of univalent category theory.}:

{
\vspace{0.15cm}
\RaggedRight
\setlength\parindent{0pt}
\setlength{\leftskip}{1cm}
\AyaKeyword{record}\hspace{0.5em}\AyaData{Precat}~\\
|\hspace{0.5em}\AyaConstructor{Ob}\hspace{0.5em}:\hspace{0.5em}\AyaKeyword{Type}~\\
|\hspace{0.5em}\AyaConstructor{Hom}\hspace{0.5em}:\hspace{0.5em}\AyaConstructor{Ob}\hspace{0.5em}\(\to\)\hspace{0.5em}\AyaConstructor{Ob}\hspace{0.5em}\(\to\)\hspace{0.5em}\AyaKeyword{Type}~\\
|\hspace{0.5em}\AyaConstructor{id}\hspace{0.5em}\((\)\AyaLocalVar{A}\hspace{0.5em}:\hspace{0.5em}\AyaConstructor{Ob}\()\)\hspace{0.5em}:\hspace{0.5em}\AyaConstructor{Hom}\hspace{0.5em}\AyaLocalVar{A}\hspace{0.5em}\AyaLocalVar{A}~\\
|\hspace{0.5em}……

\setlength{\leftskip}{0cm}
\vspace{0.15cm}
}
Suppose the syntax for instantiating records looks like:
\[
  \AyaKeyword{\mathrm{new}}~\AyaData{\mathrm{Precat}}~
  \Set{\AyaConstructor{\mathrm{Ob}}:= \cdots, \AyaConstructor{\mathrm{Hom}}:= \cdots, \cdots}
\]
In traditional sense, a \AyaKeyword{new} expression with every member specified
well-typed is a well-typed term of type \AyaData{Precat}.

Suppose we have defined the type \AyaData{Group} elsewhere with \AyaData{GroupHom}
the type of group homomorphisms. It would be nice to have these features:
\begin{itemize}
\item $\AyaData{\mathrm{Precat}} \Set{ \AyaConstructor{\mathrm{Ob}} := \AyaData{\mathrm{Group}} }$
  is the type for all instances of the record \AyaData{Precat} whose \AyaConstructor{Ob}
  member is (judgmentally equal to, similarly hereafter) \AyaData{Group}.
\item $\AyaData{\mathrm{Precat}} \Set{ \AyaConstructor{\mathrm{Ob}} := \AyaData{\mathrm{Group}},
  \AyaConstructor{\mathrm{Hom}} := \AyaData{\mathrm{GroupHom}} }$
  is the type for all instances of the record \AyaData{Precat} whose \AyaConstructor{Ob}
  member is \AyaData{Group}, and the \AyaConstructor{Hom}
  member is \AyaData{GroupHom}.
\item The same can be said about the rest of the members.
\end{itemize}
This is actually useful, because for the same carrier of objects we may have different categories,
like for the type of topological spaces $\AyaData{\mathrm{TopSpace}}$,
there is this category with morphisms being continuous maps \textsf{Top},
and there is also the category with morphisms being homotopy records of continuous maps \textsf{hTop},
and they both have type $\AyaData{\mathrm{Precat}} \Set{ \AyaConstructor{\mathrm{Ob}} := \AyaData{\mathrm{TopSpace}} }$.

To simplify the syntax, we write $\AyaData{\mathrm{Precat}} \Set{ \AyaConstructor{\mathrm{Ob}} := \AyaData{\mathrm{Group}} }$
as \AyaData{Precat} \AyaData{Group}, where the application is ordered as the record definition.

With this idea in mind, it is natural to think about \textit{definitional projection}:
\begin{itemize}
\item Suppose $A : \AyaData{\mathrm{Precat}}~\AyaData{\mathrm{Group}}$,
  then we want $A.\AyaConstructor{\mathrm{Ob}}$ to be \textit{definitionally} equal to \AyaData{Group}.
\item Suppose $A : \AyaData{\mathrm{Precat}}~\AyaData{\mathrm{Group}}~\AyaData{\mathrm{GroupHom}}$,
  then we want $A.\AyaConstructor{\mathrm{Hom}}$ to be \textit{definitionally} equal to \AyaData{GroupHom}.
\item The same can be said about the rest of the members.
\end{itemize}
This seems obvious and hence desirable, but it also touches judgmental equality,
hence affects normalization and canonicity as well.
We want more things to compute now, so certain things that are used to be normal forms
are no longer normal forms, hence we need to change the definition of normal forms,
and that complicates the metatheory the same way.

\begin{terminology}
In Arend, what is usually called \textit{records} are called \textit{classes},
except that the instances of the latter has an instance resolution mechanism similar to
type classes in Haskell.
These are unimportant for our discussion, so we will just call them records.
\end{terminology}

\begin{remark}
There is another way to solve this problem with clever $\eta$-expansion,
see \url{https://www.aya-prover.org/blog/class-defeq.html}.

This alternative approach has performance cost, which is not the case if we use extension types.
\end{remark}

\subsection{Opaque encoding for metaprogramming}
It is a long term debate that whether we should \textit{encode} the complicated types with simpler types,
or have primitive notions of complicated types.
For example, we may define natural numbers in System F as the following type:
\[
  \AyaData{\mathrm{Nat}} := \forall X. X \to (X \to X) \to X
\]
This makes the notion of inductive definitions in the type theory disappear completely.
Another example could be encoding inductive records using an inductive type with a single constructor,
as done in Lean 3~\cite{TPLean} and Lean 4~\footnote{\url{https://lean-lang.org/lean4/doc/struct.html}},
or with a $\Sigma$ type.
For instance, the definition \AyaData{Precat} above could be encoded this way:
\[
  \AyaData{\mathrm{Precat}} := \sum_{\AyaData{\mathrm{Ob}} : \AyaData{\mathrm{Type}}}
  \sum_{\AyaData{\mathrm{Hom}} : \AyaData{\mathrm{Ob}} \to \AyaData{\mathrm{Ob}} \to \AyaData{\mathrm{Type}}}
  \cdots
\]

Encoding has many advantages:
\begin{itemize}
\item They make the theory \textit{simple and expressive}: we can talk about these complicated things in a
  minimalistic type theory with just $\Pi$, $\Sigma$ types and universes,
\item They make \textit{automated theorem proving} easier, because the solvers will have fewer constructions to understand,
\item They make \textit{metaprogramming} much more pleasant, because we can now write programs that manipulate these encodings,
  which is based on a much simpler syntax. A minimalistic type theory is way easier to do
  induction (or unrestricted recursion) due to their tree-like nature\footnote{We refrain from saying they're \textit{inductive}
  or \textit{algebraic} because syntax might involve in recursions in negative positions, which is not allowed in inductive types.},
  while the primitive inductive types have a way more complicated inherent structure.
\end{itemize}

They also have disadvantages:
\begin{itemize}
\item \textit{Error messages} will be unreadable. There are times we want to talk about theorems on natural numbers,
  lists, or trees, but somehow due to the encoding they can become theorems about long functions,
  and it is easier to get lost,
\item \textit{Interactive theorem proving} becomes painful. What constructor should I apply?
  How should I think about this data structure?
  It is the type that should give this kind of information, but now it's gone,
\item Without the encoding, type checking is more \textit{efficient}. Comparing two long function types is less efficient than just
  comparing two names.
\end{itemize}

It would be so nice if we can work with primitive inductive types or records most of the time,
and whenever we need to talk about metatheory, do metaprogramming, or invoke a solver,
we replace with encoding.
If at the end of the day everything is encoded, compiling the language would also be easier.

But how do we control this? This sounds a bit like controlling unfolding,
but instead of definitions, we talk about unfolding types themselves.
Looks like they can be solved by the same mechanism!






\section{Introduction to extension types}
Suppose we are working on a type theory based on Martin-L\"of's dependent type theory~\cite{MLTT}.
This section is devoted to decipher the following things:
\[
  \Set{A|φ ▷ u}
  \qquad
  Γ,φ ⊢ u:A
\]
\subsection{Context restrictions with propositions}
We add a new ``universe of \textit{strict} propositions'' \AyaData{F} to our type theory,
which we think of as a type, and instances as certain \textit{propositions}.
They are called so because we wish to talk about the \textit{truthness} of these instances.

This is not to be confused with the ``propositions-as(-some)-types'' propositions.
For instance, if there were to have logical connectives in \AyaData{F},
they will be unrelated to the formation rules in the type theory.
We will not have implication in \AyaData{F} just because we have function type.
\begin{remark}[ST$λ$C]\label{rem:prop-syntax}
From a syntax perspective of type theory,
the treatment of \AyaData{F} compared to the universe of propositions as in Curry--Howard correspondence,
is similar to how Russell's simple theory of types~\cite{RussellSTT} is related to simply typed $λ$-calculus:
The instances of \AyaData{F} do not correspond to types, nor do they correspond to \textit{type codes}
as in universe \'a la Tarski~\cite{IndRec}. Instead, they are just terms, and the logical deduction on these terms
follows its own system, and is nothing to do with things like function types in the type theory.

This is also mentioned in~\cite[Remark 2]{CU} in a different context.
\end{remark}
\begin{remark}\label{rem:prop-nltt}
Actually,~\cref{rem:prop-syntax} is no longer true if we are to think about semantics
or type theories with multiple layers, because these propositions can be thought of as types
from another \textit{layer}.
But we will avoid those discussions to keep the focus
on the syntax and implementation aspects of type theories.
\end{remark}

We will keep working with the universe of propositions abstractly when
defining the syntax, and then talk about the intuition and examples.
If readers have trouble understanding the syntax, they are encouraged to jump between the latter
sections and contents below to avoid brain stack overflow.

With \AyaData{F}, we introduce the notion of \textit{context restriction}, which gives rise to judgments of this form,
assuming \fbox{$Γ⊢ φ :\AyaData{F}$} and $\mathcal J$ denotes a judgment such as \fbox{$u:A$},
\fbox{$u≡ v:A$} and things like that:
\[
  Γ, φ ⊢ \mathcal J
\]
The idea is that we add $φ$ to $Γ$ to get a new context,
we call it \textit{restricting} the context by $φ$.
Then, it is part of the \textit{assumptions},
which is why we write it at the left-hand side of the $⊢$.
When substituting over this judgment, we not only need to provide an instantiation for $Γ$,
but also a proof for $φ$.
Naturally, we assume $φ$ is true when type checking $\mathcal J$.
\begin{observation}
The structure of the propositions in \AyaData{F} has to be very simple,
because they affect typing.
It is very easy to break decidability of type checking
if we don't pay attention on restricting these propositions.
\end{observation}
\begin{example}
Suppose we can talk about judgmental equality in $\AyaData{F}$,
then we essentially get equality reflection,
because we can now say things like
(where $u ≡ v$ is a term of type $\AyaData{F}$):
\[
  Γ, a: P(u), u ≡ v ⊢ a: P(v)
\]
And this judgment should hold. Essentially we have a way to
have general judgmental equality assumptions.
This is both an example of how to think about context restrictions in action
and how to abuse them to break decidability.
\end{example}

\begin{remark}[Semantics]\label{rem:prop-sem}
Continuing the discussion in~\cref{rem:prop-nltt}: from a semantic point of view,
judgments with propositions in the context can be thought of as really passing their proofs.
But this all happens in the meta-level,
and the proofs need to be done automatically in the type checker.
\end{remark}

\subsection{Extension types and their typing rules}
Context restriction is the tool we need to define extension types.

The rules (in the order of formation, introduction, elimination, computation)
of extension types are written in~\cref{fig:extty}.
\begin{exercise}
Define the uniqueness rule.
\end{exercise}

\begin{figure}[h!]
\begin{mathpar}
\inferrule
  { \Gvdash \isType A \\ \Gvdash φ : \AyaData{F} \\
    Γ,φ ⊢ u : A }
  { \Gvdash \isType{\Set{A | \overline{φ ▷ u}}} }
\and
\inferrule
  { \Gvdash v : A \\ \overline{Γ,φ ⊢ u ≡ v : A} }
  { \Gvdash \inS{v} : \Set{A | \overline{φ ▷ u}} }
\and
\inferrule
  { \Gvdash v : \Set{A | \overline{φ ▷ u}} }
  { \Gvdash \outS{v} : A \\ \text{and} \\
    \overline{Γ,φ⊢ \outS{v} ≡ u : A} }
\and
\inferrule{\Gvdash v : A}{\Gvdash \outS{\inS{v}} ≡ v : A}
\end{mathpar}
\caption{Rules of extension types}
\label{fig:extty}
\end{figure}

Let's read the rules one by one.
\begin{itemize}
\item The formation rule says, an extension type is like a wrapper of a given type $A$,
  and a term $u$ that is of type $A$ when $φ$ is true.
  The $φ ▷ u$ part is called a \textit{clause} can appear multiple times.
\item The introduction rule says, given a term $v:A$, if it is judgmentally equal to $u$ when $φ$ is true,
  then we can turn it into a term of the extension type.
  In case there are multiple clauses, they all need to be satisfied.
\item The elimination rule is the inverse of the introduction: we can unwrap the extension type,
  and the unwrapper computes (without knowing what is inside) to $v$ if $φ$ is true.
  This is guaranteed in the introduction rule.
\item The computation rule says introduction followed by elimination is doing nothing. 
\end{itemize}
These rules make essential use of context restrictions,
and it behaves like \textit{subtype} of the original type where only terms that satisfy this
``judgmental equality condition'' is in this subtype.
Then, let's see how to apply this to solve all the problems in~\cref{sec:problems}.
For every problem, we start by saying what is the corresponding \AyaData{F}.

\begin{remark}[Semantics]\label{rem:extt}
Recall~\cref{rem:prop-sem}. We can make the analogy between $\Set{A | φ ▷ u}$
and the set-builder notation $\Set{x \in A | φ → (x ≡ u : A) }$
(where we think of $φ$ as the type of its proofs),
i.e. a \emph{subtype} of $A$ where given a proof of the proposition $φ$,
we give you back the proof that $x ≡ u$.
This justifies all the other rules.
\end{remark}

\begin{remark}\label{rem:extent}
An alternative name for extension types is \textit{extent types} as used in~\cite[\S 3.5*5]{JS}.
Intuitively, the extension type \textit{constraints} or \textit{restricts} the original type,
as opposed to \textit{extending} it, which might cause some confusion. The word \textit{extent}
works nicely as ``the extent to which the formula holds''.

A similar idea can be found in~\cite[\S 1.3.3, \S 2.1.6]{LRType},
which uses the name \textit{static extent}.
\end{remark}

\section{Extension types in action}
\subsection{Controlling unfolding and metaprogramming}
For controlling unfolding, we for every definition $f$ create a corresponding proposition
$φ_f : \AyaData F$, meaning that ``we want to unfold $f$''.
Then, this is what happens when we call functions:
\begin{itemize}
\item Now, whenever we say in the context ``we want to unfold $f$'',
  we add $φ_f$ to the context and restrict the context by it.
\item Every definition $f$ gets renamed to $\delta_f$,
  and references to it gets wrapped into an extension type:
  \[
    f : \Set{A | φ_f ▷ \delta_f}
  \]
  So, it only unfolds to its actual definition when we say so.
\item When we call a function, say, write $f$ in the source code,
  the type checker translate that into $\outS f$.
  This term has the same type as in the source code,
  and the only difference is in judgmental equality.
\end{itemize}
For every definition $f$, it declares which other definitions it unfolds (say, $g$),
and this is used to restrict the context when type checking $f$,
and in the logic of \AyaData F, we add a rule $φ_f \implies φ_g$.
This makes unfolding automatically transitive due to the transitive nature of logical implication.
\begin{notation}
In the original paper~\cite{CU}, the proposition is called $\mathbf{\Upsilon}_f$,
and in their HoTT 2023 abstract\footnote{
  \url{https://hott.github.io/HoTT-2023/abstracts/HoTT-2023_abstract_19.pdf}},
it's called $φ_f$.
\end{notation}
Then we get the following guarantee:
\begin{itemize}
\item Unfolding is totally controlled by the context restriction, which is controlled by the user.
\item Unfolding is transitive and only transitive ---
  say I call $f$ and $g$ separately, and $f$ unfolds $g$, and I want to unfold $f$.
  My other call to $g$ will not be affected.
\item Modifications of judgmental equality is achieved with extension types,
  whose metatheory is well understood.
\end{itemize}

\begin{remark}\label{rem:singl}
Controlling unfolding uses extension types as if it's a singleton type~\cite{Singletons},
because there is only one \textit{clause} in every extension type.
In fact, everything can be achieved with singletons can be achieved with extension types
in a similar way. 
\end{remark}

We also add a constant $\text{\faFolderOpenO{}} : \AyaData F$,
which says we want to replace types with their encoding.
Then, without \faFolderOpenO{} in the context, types look primitive,
and when we need to work with their encoding, we add \faFolderOpenO{} to the context.

\begin{remark}
The constant \faFolderOpenO{} is very similar to the $\text{\faUnlock{}}_{\mathsf{st}}$ symbol
(both as an operator on context and a symbol in the extent type) in~\cite[\S 1.3.3, \S 2.1.6]{LRType},
where they only need a single constant for their purpose.
So they don't need a dedicated type \AyaData{F}.
\end{remark}

If every metaprogram automatically declare \faFolderOpenO{},
the reflected syntax tree can be encoding only,
and when compiling the program, we set \faFolderOpenO{} to true,
then all the complicated type definitions will be gone.

\subsection{Definitional projection with singletons}
The high-level idea is that we use extension types as singletons (see~\cref{rem:singl}),
which requires creating a new record type for every \textit{partially specified record}.

In \cooltt, this mechanism is called \textit{record patching}, which replaces fields with
extension types that specify the expected value on that field.
It requires structural records to work.

\begin{example}
Consider the following example:

{
\vspace{0.15cm}
\RaggedRight
\setlength\parindent{0pt}
\setlength{\leftskip}{1cm}
\AyaKeyword{record}\hspace{0.5em}\AyaData{Sing}~\\
|\hspace{0.5em}\AyaConstructor{A}\hspace{0.5em}:\hspace{0.5em}\AyaKeyword{Type}~\\
|\hspace{0.5em}\AyaConstructor{a}\hspace{0.5em}:\hspace{0.5em}\AyaConstructor{A}

\setlength{\leftskip}{0cm}
\vspace{0.15cm}
}

When we write $\AyaData{\mathrm{Sing}}~T$ for a known type $T$,
it gets translated into the following new record type,
where $1:\AyaData{F}$ denotes the truth proposition:

{
\vspace{0.15cm}
\RaggedRight
\setlength\parindent{0pt}
\setlength{\leftskip}{1cm}
\AyaKeyword{record}\hspace{0.5em}\AyaData{Sing}~\\
|\hspace{0.5em}\AyaConstructor{A}\hspace{0.5em}:\hspace{0.5em}\(\{\)\hspace{0.5em}\AyaKeyword{Type}\hspace{0.5em}|\hspace{0.5em}1\hspace{0.5em}↦\hspace{0.5em}\AyaLocalVar{T}\hspace{0.5em}\(\}\)~\\
|\hspace{0.5em}\AyaConstructor{a}\hspace{0.5em}:\hspace{0.5em}\AyaKeyword{outS}\((\)\AyaConstructor{A}\()\)

\setlength{\leftskip}{0cm}
\vspace{0.15cm}
}

It enforces $A$ to be $T$, and projection of $A$ will compute to $T$ unconditionally
due to the computation rule of extension types.
\end{example}

\subsection{Definitional projection with projective extension types}\label{sub:defproj}
Another approach is to use a new extension type designed specifically for records
and their projections.
For record $R$ with field $f$ specified to be $u$, we write:
\[
  \Set{R | f := u}
\]
In case of more fields, we put multiple clauses.
Note that this implies for every field $f$,
we create an instance of \AyaData{F} under the same name.

The example in~\cref{sub:motive-defproj} will be translated like this:
\begin{align*}
  \AyaData{\mathrm{Precat}} \Set{ \AyaConstructor{\mathrm{Ob}} := \AyaData{\mathrm{Group}} }
    &\implies \Set{\AyaData{\mathrm{Precat}} | \AyaConstructor{\mathrm{Ob}} ▷ \AyaData{\mathrm{Group}} } \\
  \AyaData{\mathrm{Precat}} \Set{ \begin{array}{ll}
    \AyaConstructor{\mathrm{Ob}} &:= \AyaData{\mathrm{Group}}\\
    \AyaConstructor{\mathrm{Hom}} &:= \AyaData{\mathrm{GroupHom}}
  \end{array} } &\implies
  \Set{\AyaData{\mathrm{Precat}} | \begin{array}{ll}
    \AyaConstructor{\mathrm{Ob}} &▷ \AyaData{\mathrm{Group}} \\
    \AyaConstructor{\mathrm{Hom}} &▷ \AyaData{\mathrm{GroupHom}}
  \end{array} }
\end{align*}

Now we define the rules of this new extension type,
called \textit{projective extension type},
with the syntax of records treated informally:

\begin{mathpar}
\inferrule{\Gvdash R~\text{is \AyaKeyword{record}} \\ \Gvdash \overline{f~\text{is a field of type}~T~\text{of}~R} \\
  Γ ⊢ \overline{u : T}}{\Gvdash \isType{\Set{R | \overline{f ▷ u}}}} \and
\inferrule{\Gvdash \overline{r.f ≡ u : T}}{\Gvdash \inS{r} : \Set{R | \overline{f ▷ u}}} \and
\inferrule{\Gvdash r : \Set{R | \overline{f ▷ u}}}{
  \Gvdash \outS{r} : R \\
  \text{and} \\
  \Gvdash \overline{r.f ≡ u : T}
} \and
\inferrule{\Gvdash r : R}{\outS{\inS r} ≡ r : R}
\end{mathpar}

Compared to the rules of extension types, the only differences are:
\begin{itemize}
\item For clause $f ▷ u$, instead of having $u:R$ like in extension types,
  we let $u:T$ where $T$ is the type of $f$,
\item We additionally allow projections on the extension type,
  and they compute to the expected value.
\end{itemize}

Note that in the rules presented above, we also treated the iterative nature of record fields informally.
In fact, for every specified field, we need to have that known type available in the type checking of the
rest of the clauses. This is because records are dependent, and some clauses will only make sense when
other clauses are specified.

\begin{remark}
There could possibily be a dual of projective extension types,
called \textit{coprojective extension types}.
Since projective extension types add judgmental constraints to the introduction rule of records,
coprojective extension types add judgmental constraints to the elimination rule of inductive types -- pattern matching.

A potential application of this is discussed in the author's presentation
in the Workshop on the Implementation of Type Systems\footnote{\url{https://www.youtube.com/watch?v=I7r4YQOU0Ws}}.
\end{remark}

\subsection{Type theory with shapes}
In cubical type theory, \AyaData{F} corresponds to certain constraints on variables of the interval type $\mathbb{I}$,
for instance $(i=0)$ for $i:\mathbb{I}$ is a well-typed term of type \AyaData{F}.
The notation here might be a bit confusing, here's the typing rule version:
\[
  i:\mathbb{I} ⊢ (i=0) : \AyaData{F}
\]
The $=$ and $0$ in this expression are not the usual equality and zero,
instead they are just keywords.
We may use the following type to encode the path type $a =_A b$:
\[
(i : \mathbb I) \to \Set{A | (i = 0) ▷ a, (i = 1) ▷ b}
\]
Elements of this type will behave exactly the same as paths in cubical type theory,
where applied by $0$ it returns the left endpoint $a$, and applied by $1$ it returns the right endpoint $b$.
Introduction of this type behaves exactly the same as the introduction of paths, no surprise.

Note that this is exactly what is called an extension type in cubical type theory,
where the free variables in the propositions $φ$ are all bound in the extension type.
In the PL applications, all propositions are constants,
so we don't need to have a place for all those binders.
The unbounded version of this in cubical type theory is called a \emph{cubical subtype}.

\begin{remark}
If we think of the $\mathbb I$ and its products as a type theory
(separately from the dependent type theory),
then these propositions constitute a \emph{logic over} that type theory.
\end{remark}
The formation rule for extension types in cubical looks like the following:

\begin{mathpar}
\inferrule
  { Φ,\overline{i:\mathbb I} ⊢ φ : \AyaData{F} \\
    Φ,\overline{i:\mathbb I}; Γ, φ ⊢ \isType A \\
    Φ,\overline{i:\mathbb I}; Γ, φ ⊢ u:A }
  { Φ,Γ ⊢ [\overline{i:\mathbb I}]\isType{\Set{A | \overline{φ ▷ u}}} }
\end{mathpar}

The context is \emph{dyadic}, that is to say,
split into ``interval context'' $Φ$ and the usual ``type theory context'' $Γ$,
and all the terms in the formation are bound in the interval variables in the list $\overline{i:\mathbb I}$.

For more information and why is this more desirable compared to the usual path type, see~\cite{DMCubical}.
We sketch some of the ideas here:
\begin{itemize}
\item It has the ability to express path composition (transitivity of propositional equality in traditional language)
  without using implicit arguments on every endpoint, and similarly for other lemmas on paths. Pseudocode:

  {
\vspace{0.15cm}
\RaggedRight
\setlength\parindent{0pt}
\setlength{\leftskip}{1cm}
\AyaKeyword{def}\hspace{0.5em}\AyaFn{comp}\hspace{0.5em}:\hspace{0.5em}\(\{\)\AyaLocalVar{A}\hspace{0.5em}:\hspace{0.5em}\AyaKeyword{Type}\(\}\)\hspace{0.5em}\((\)\AyaLocalVar{p}\hspace{0.5em}:\hspace{0.5em}[\hspace{0.5em}\AyaLocalVar{i}\hspace{0.5em}]\hspace{0.5em}\(\{\)\hspace{0.5em}\AyaLocalVar{A}\hspace{0.5em}\(\mid\)\hspace{0.5em}\(\}\)\()\)~\\
\hspace{1.0em}\hspace{1.0em}\((\)\AyaLocalVar{q}\hspace{0.5em}:\hspace{0.5em}[\hspace{0.5em}\AyaLocalVar{i}\hspace{0.5em}]\hspace{0.5em}\(\{\)\hspace{0.5em}\AyaLocalVar{A}\hspace{0.5em}\(\mid\)\hspace{0.5em}\((\)\AyaLocalVar{i}\hspace{0.5em}=\hspace{0.5em}0\()\)\hspace{0.5em}▷\hspace{0.5em}\AyaLocalVar{p}\hspace{0.5em}1\hspace{0.5em}\(\}\)\()\)~\\
\hspace{1.0em}\(\to\)\hspace{0.5em}[\hspace{0.5em}\AyaLocalVar{i}\hspace{0.5em}]\hspace{0.5em}\(\{\)\hspace{0.5em}\AyaLocalVar{A}\hspace{0.5em}\(\mid\)\hspace{0.5em}\((\)\AyaLocalVar{i}\hspace{0.5em}=\hspace{0.5em}0\()\)\hspace{0.5em}▷\hspace{0.5em}\AyaLocalVar{p}\hspace{0.5em}0\hspace{0.5em}\(\mid\)\hspace{0.5em}\((\)\AyaLocalVar{i}\hspace{0.5em}=\hspace{0.5em}1\()\)\hspace{0.5em}▷\hspace{0.5em}\AyaLocalVar{q}\hspace{0.5em}1\hspace{0.5em}\(\}\)

\setlength{\leftskip}{0cm}
\vspace{0.15cm}
}

  This saves the burden of unification in the type checker.
\item It makes working with higher-dimensional paths easier, because they'll just be functions with multiple parameters
  and having regular boundaries. Expressing these paths using the usual path type is a bit more painful.
\end{itemize}

In simplicial type theory where extension types are first introduced,
the definition is much more general and fine-grained.
It starts with a general notion of \emph{type theory with shapes} where
there is the type of \emph{cubes}, e.g. an interval type $\mathbb I$ and its products,
and the type of \emph{shapes} described as propositions on cubes.
So, saying $\overline{i:\mathbb I}⊢φ:\AyaData F$ is similar to saying ``$\Set{\overline{i}|φ}$ is a shape'',
and $φ$ by itself is referred to as a \emph{tope}.
\begin{example}
In the context $i,j:\mathbb I$, the truth proposition is the full square,
the proposition $(i=0)$ is an edge of the square,
and proposition $(i≤j)$ is a triangle in the square.
\end{example}
Based on this idea, we first rewrite the cubical version of extension type using
the notation in simplicial type theory~\cite[\S 2.2]{InfCat}:

\begin{mathpar}
\inferrule
  { Φ,\overline{i:\mathbb I} ⊢ φ~\text{shape} \\
    Φ,\overline{i:\mathbb I}; Γ, φ ⊢ \isType A \\
    Φ,\overline{i:\mathbb I}; Γ, φ ⊢ u:A }
  { Φ,Γ ⊢ \isType{\Braket{\prod\nolimits_{\overline{i:\mathbb I}} A |^φ_u}} }
\end{mathpar}

Then, we allow $A$ and $u$ to be in difference shapes.
This gives us the very original definition of extension types:

\begin{mathpar}
\inferrule
  { Φ,\overline{i:\mathbb I} ⊢ φ~\text{shape} \\
    Φ,\overline{i:\mathbb I} ⊢ θ~\text{shape} \\
    Φ,\overline{i:\mathbb I}, φ ⊢ θ \\
    Φ,\overline{i:\mathbb I}; Γ, θ ⊢ \isType A \\
    Φ,\overline{i:\mathbb I}; Γ, φ ⊢ u:A }
  { Φ,Γ ⊢ \isType{\Braket{\prod\nolimits_{\overline{i:\mathbb I}~\mid~θ} A |^φ_u}} }
\end{mathpar}

\section{Related Work}
Some places I've seen extension types are:
\begin{itemize}
\item Simplicial type theory by Riehl and Shulman~\cite{InfCat}, which first introduced the idea of extension types,
\item The \textit{nLab} entry on \textit{type theories with shapes}\footnote{\url{https://ncatlab.org/nlab/show/type+theory+with+shapes}}
  together with the entry on \textit{extension types}\footnote{\url{https://ncatlab.org/nlab/show/extension+type}},
\item The PhD thesis of Jonathan Sterling~\cite[\S 3.5]{JS},
\item Controlling unfolding in type theory by Daniel Gratzer \textit{et al.}~\cite[\S 3]{CU}.
\item Logical relations as types by Jonathan Sterling \textit{et al.}~\cite{LRType}.
\end{itemize}

\printbibliography
\end{document}